# Comment on "Calculation of the Masses of All Fundamental Elementary Particles with an Accuracy of Approx. 1%" [Journal of Modern Physics, 2010, 1, 300-302]

Adrian Radu

*Department of Physics, "Politehnica" University of Bucharest, RO-77206 Bucharest, Romania*

In a recent paper [1], K. O. Greulich propose an empirical, one-parameter rule (Eq.(1) in the original reference) which is claimed to predict the masses of all fundamental elementary particles:

$$m/m_{electron} = N/2\alpha. \qquad (1)$$

This statement seems to be true as sustained by the comparison of some experimental and calculated elementary masses (Table 2 in the original reference). The author admits that in the case of an isolated single particle a good result could be achieved simply by chance. He also states that the probability to fit by chance the whole ensemble of all elementary masses within 1% accuracy is close to zero. However, one may observe that only 8 of the 11 particles which are relevant for the correctness of the Eq.(1) satisfy the formula within 1% accuracy and this result could have been achieved simple by chance. We will demonstrate these assertions in two steps.

1. **It doesn't make sense to sustain the validity of Eq.(1) by using particles with $N \geq 50$.**

Let's put Eq.(1) in the equivalent forms $m_{calc}^N = N \dfrac{m_{electron}}{2\alpha} = N \Delta m \cong N \cdot 35.01\, MeV/c^2$; one may directly verify that $\forall\, m_{exp} \geq 50\Delta m \cong 1750.63\, MeV/c^2$, $\exists\, N \in \mathbb{N}^*$, $N \geq 50$, s.t. $\left|m_{exp} - m_{calc}^N\right|/m_{calc}^N < 1\%$, which means that for any particle heavier than $50\Delta m$ a fit accurate within 1% may always be found. The author himself states that "an accurate calculation via Equation (1) of heavy masses is trivial" and talks about the other 11 particles, but he still compares in Table 2 all the particles' experimental/calculated masses. This over-extended comparison could be interpreted as a misleading presentation.

For the next step of our discussion, we eliminate the 10 non-relevant particles with $m_{exp} \geq 50\Delta m$. There are 3 of the remaining 11 particles (mesons Eta, Omega, baryon Xi) whose masses don't fit Eq.(1) within 1% accuracy.

2. **The baryons and even the mesons could have satisfied Eq.(1) simple by chance.**

If we take $N_1 + 1 < N_2 \leq 50$ and without any prior information on the elementary masses, the probability to achieve a good result simply by chance for a single particle with $N_1 \Delta m \leq m_{exp} \leq N_2 \Delta m$ can be expressed as

$$p_{N_1-N_2} = \left[1\%\, N_1 + 2\%\, \sum_{k=N_1+1}^{N_2-1} k + 1\%\, N_2\right]/(N_2 - N_1) = (N_1 + N_2)\%. \qquad (2)$$

The probability that within a group of G particles with $N_1\Delta m \leq m_{exp} \leq N_2\Delta m$, S experimental masses verify Eq.(1) within 1% accuracy simple by chance is given by the binomial formula

$$P_G^S = \frac{G!}{S!(G-S)!}\, p_{N_1-N_2}^S \left(1 - p_{N_1-N_2}\right)^{G-S}. \qquad (3)$$

Let's now use Eqs.(2,3) to calculate the probability that 4 of the relevant 5 baryons in the original Table 2 satisfy Eq.(1) only by chance. According to the table, we must take $N_1 = 26$ and $N_2 = 48$, which leads to $p_{26-48} = 74\%$ and furthermore to $P_5^4 \cong 38.98\%$. It's quite probable!

To calculate the probability that 7 particles of 10 (relevant mesons and baryons) satisfy Eq.(1) simple by chance we must take $N_1 = 3$ and $N_2 = 48$, which leads to $p_{3-48} = 51\%$ and furthermore to $P_{10}^7 \cong 12.67\%$. Even if this probability is smaller, by no means this could be considered a probability close to zero.

If K. O. Greulich had argued on how small a "by chance-fit probability" must be for validating Eq.(1), it would have been acceptable that his model correctly describes some particles with smaller masses, like the Muon, Pion and Kaon. Indeed, for $N_1 = 3$ and $N_2 = 15$, one may obtain $p_{3-15} = 18\%$ and $P_3^3 \cong 0.58\%$.